\documentclass[usenatbib]{nature_djm}

\pdfoutput=1

\usepackage{balance}
\usepackage{graphicx}
\usepackage{widetext}

\newlength{\figwidth}
\newlength{\thinfigwidth}
\newlength{\tinyfigwidth}
\newlength{\widefigwidth}
\def\HI{H\,{\sc i}}
\def\HII{H\,{\sc ii}}

\def\la{\ifmmode\stackrel{<}{_{\sim}}\else$\stackrel{<}{_{\sim}}$\fi}
\def\ga{\ifmmode\stackrel{>}{_{\sim}}\else$\stackrel{>}{_{\sim}}$\fi}

\def\fdg{\hbox{$.\!\!^\circ$}}
\def\mathbi#1{\textbf{\em #1}}

\pubdate{13 OCTOBER 2011}

\title[ ]{Low-Mach-number turbulence in interstellar gas \\ revealed 
by radio polarization gradients}

\author[ ]
{B. M. Gaensler,$^1$
M.~Haverkorn,$^{2,3,4}$
B.~Burkhart,$^5$ 
K.~J.~Newton-McGee,$^{1,6}$ 
\newauthor
R.~D.~Ekers,$^6$ 
A.~Lazarian,$^5$
N.~M.~McClure-Griffiths,$^6$ 
T.~Robishaw,$^1$
\newauthor
J.~M.~Dickey$^7$ \& A.~J.~Green$^1$}

\begin{document}

\maketitle

\newcommand{\affiliations}
{
\begin{widetext}
\small{\textsf{
\hspace*{-8mm}
$^{1}$Sydney Institute for Astronomy, School of Physics, The University
of Sydney, NSW 2006, Australia.
$^2$ASTRON, Oude Hoogeveensedijk 4, 7991 PD Dwingeloo, The Netherlands.
$^3$Leiden Observatory, Leiden University, P.O. Box 9513, 2300 RA Leiden,
The Netherlands.
$^4$Department of Astrophysics/IMAPP, Radboud University Nijmegen, PO Box
9010, 6500 GL Nijmegen, The Netherlands.
$^5$Astronomy Department, University of Wisconsin, Madison, 475 North
Charter Street, Madison, WI 53711, USA. 
$^6$Australia Telescope National Facility, CSIRO Astronomy and Space
Science, PO Box 76, Epping, NSW 1710, Australia.
$^7$School of Mathematics and Physics, University of Tasmania, Private Bag 37,
Hobart, TAS 7001, Australia.
}
}
\end{widetext}

}

\maketitle

\noindent
{\textbf{The interstellar medium of the Milky Way is 
multi-phase,\cite{fer01} magnetized\cite{db05b} and turbulent.\cite{cl10}
Turbulence in the interstellar medium produces a global cascade of
random gas motions, spanning scales ranging from 100~parsecs to
1000~kilometres.\cite{ars95} Fundamental parameters of
interstellar turbulence such as
the sonic Mach number (the speed of sound) have been difficult to determine because observations
have lacked the sensitivity and resolution to directly image the small-scale
structure associated with turbulent motion.\cite{knpw07,klb07,kkfg08}
Observations of linear polarization and Faraday rotation in radio emission
from the Milky Way have identified unusual polarized structures that often
have no counterparts in the total radiation intensity or at other
wavelengths,\cite{wdj+93,gldt98,hkd00, gdm+01,ul02} and whose physical
significance has been unclear.\cite{sb03,hh04b,fs07} Here we report that the
gradient of the Stokes vector ({\em Q}, {\em U}), where {\em Q}\ and {\em
U}\ are parameters describing the polarization state of radiation, provides an image of
magnetized turbulence in diffuse ionized gas, manifested as a complex
filamentary web of discontinuities in gas density and magnetic field.
Through comparison with simulations, we demonstrate that turbulence in the
warm ionized medium has a relatively low sonic Mach number, $\mathcal{M}_s
\la$~2.  The development of statistical tools for the analysis of polarization
gradients will allow accurate determinations of the Mach number, Reynolds
number and magnetic field strength in interstellar turbulence over a wide range
of conditions.}}

We consider radio continuum images of an 18-square-degree
patch\cite{gdm+01,mgd+01} of the Galactic plane, observed with the Australia
Telescope Compact Array (ATCA) at a frequency of 1.4~GHz.  Data were
simultaneously recorded in total intensity (Stokes parameter~$I$) and in
linear polarization (Stokes parameters~$Q$ and $U$).  The Stokes~$I$ image
(Fig.~1) shows a typical distribution of radio emission, consisting of
supernova-remnant shells, ionized regions around massive stars (\HII\
regions), and unresolved distant radio sources.  However, the corresponding
images of $Q$, $U$, and the linearly polarized intensity $P \equiv
(Q^2+U^2)^{1/2}$ in Figure~1 are filled with complex structure that bears
little resemblance to the Stokes~$I$ image, as has also been seen in many
other polarimetric observations at radio
frequencies.\cite{wdj+93,gldt98,ul02} The intensity variations seen in $Q$,
$U$ and $P$ are the result of small-scale angular structure in the Faraday
rotation induced by ionized gas,\cite{wdj+93} and are thus an indirect
representation of turbulent fluctuations in the free-electron density and
magnetic field throughout the interstellar medium\cite{fs07} (ISM).

\begin{figure*}
\includegraphics[width=\textwidth]{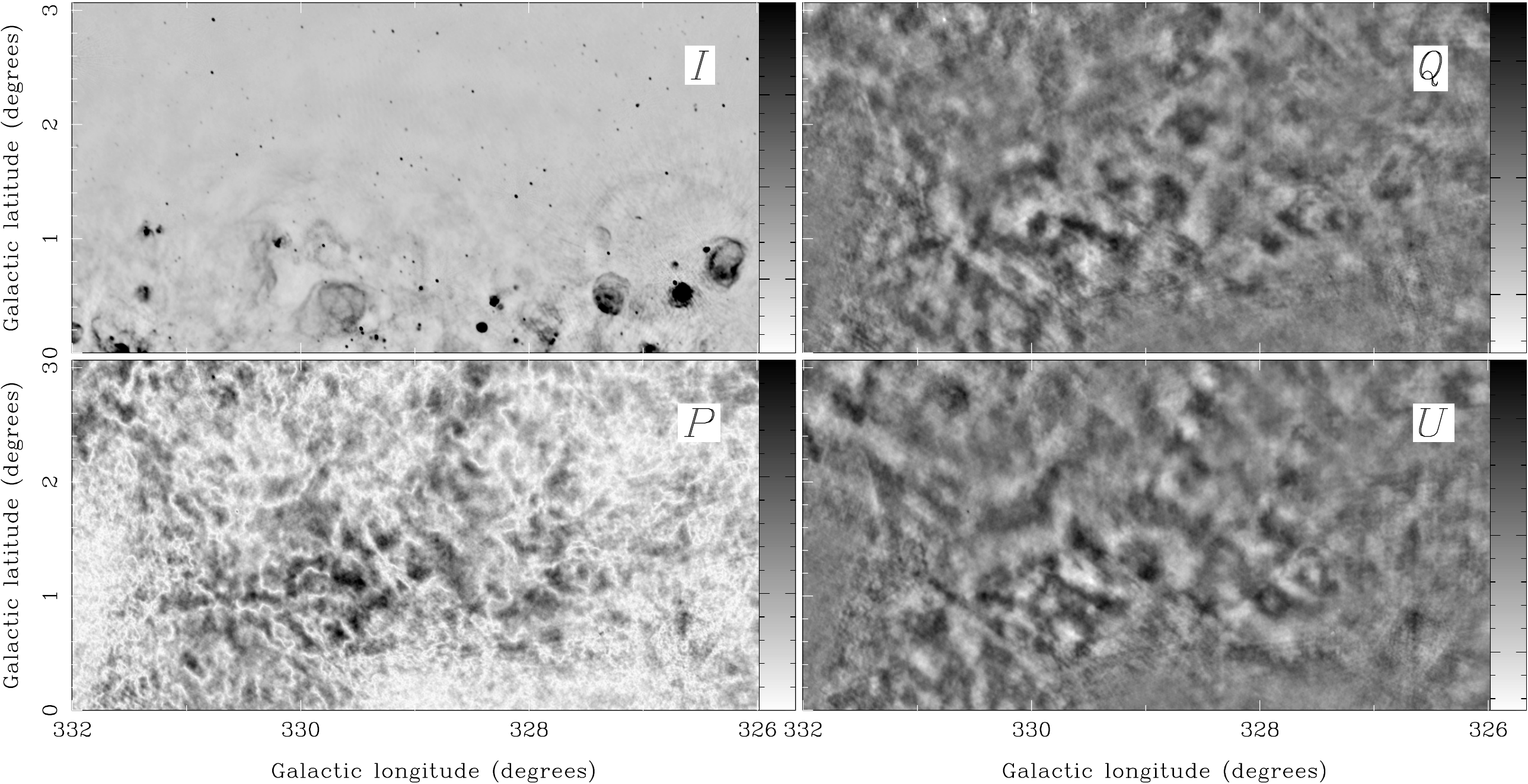}
\caption{$|$ {\bf{Total intensity ($\mathbi{I}$) and linearly polarized
intensity ($\mathbi{Q}$, $\mathbi{U}$, $\mathbi{P}$)
for an 18-deg$^2$ region of the
Southern Galactic Plane Survey.\protect\cite{hgm+06}}} All four images have been
generated\protect\cite{gdm+01} from a set of
observations\protect\cite{mgd+01} taken
at the ATCA over the period 1997~April to
1998~April using a 96-MHz bandwidth centred on an observing frequency of
1384~MHz.  The field is a mosaic of 190 pointings each with a total
integration time of 20~min, resulting in an approximately uniform
sensitivity over most of the field, of 0.8~mJy per beam (Stokes~$I$)
or 0.55~mJy per beam (Stokes~$Q$ \& $U$) at an angular resolution
of 75~arcseconds (1~Jy~$\equiv10^{-26}$~W~m$^{-2}$~Hz$^{-1}$).  The
scale for each image is shown on the right of each panel.
The Stokes~$I$ image is displayed over a range of --40 to +150~mJy
per beam.  Because the ATCA is an interferometer, it is not sensitive to
structure on angular scales larger than 35~arcminutes.  Faint wisps can be
seen, corresponding to the sharp edges of large-scale structures.  However,
the bulk of the smooth radio emission from Galactic cosmic rays is not
detected.  Imaging artefacts in the form of grating rings and radial streaks
can be seen around a few very bright sources, but these regions were not
used in our statistical analysis.  The Stokes~$Q$ and $U$ images are
displayed over a range of --15 to +15~mJy per beam, and the $P$
image covers a range of 0 to 15~mJy per beam. Almost none of the
structure seen in $Q$, $U$ and $P$ has any correspondence with any emission
seen in Stokes~$I$; the mottled structure results from spatial fluctuations
in Faraday rotation in the ISM.}
\label{fig_iqup}
\end{figure*}

A limitation of previous studies is that they usually interpreted the data
in terms of the amplitude, $P$, and/or angle, $\theta \equiv \frac{1}{2}
\tan^{-1} (U/Q)$, of the complex Stokes vector $\mathbi{P} \equiv (Q,U)$.
However, neither polarization amplitude nor polarization angle is preserved
under arbitrary translations and rotations in the  $Q-U$ plane. These can
result from one or more of a smooth distribution of intervening polarized
emission, a uniform screen of foreground Faraday rotation, and the effects
of missing large-scale structure in an interferometric data set. In the most
general case, we are thus forced to conclude that the observed values of $P$
and $\theta$ do not have any physical significance, and that only
measurements of quantities that are both translationally and rotationally
invariant in the $Q-U$ plane can provide insight into the physical
conditions that produce the observed polarization distribution.

The simplest such quantity is the spatial gradient of $\mathbi{P}$, i.e.,
the rate at which the polarization vector traces out a trajectory in the
$Q-U$ plane as a function of position on the sky. The magnitude of the
gradient is unaffected by rotation and translation, and so has the potential
to reveal properties of the polarization distribution that might otherwise
be hidden by excess foreground emission or Faraday rotation, or in data sets
from which large-scale structure is missing (as is the case for the data
shown in Fig.~1).  The magnitude of the polarization gradient is:
\smallskip
\begin{equation}
\hspace{5mm} |\nabla \mathbi{P}| = \sqrt{\left(\frac{\partial Q}{\partial x}\right)^2
+ \left(\frac{\partial U}{\partial x}\right)^2
+ \left(\frac{\partial Q}{\partial y}\right)^2
+ \left(\frac{\partial U}{\partial y}\right)^2} .
\label{eqn_gradp}
\end{equation}
The expression in Equation~(\ref{eqn_gradp}) can be  calculated simply,
and the corresponding image of $|\nabla \mathbi{P}|$ (Fig.~2) reveals a
complex network of tangled filaments.  In particular, all
regions in which $|\nabla \mathbi{P}|$ is high consist of elongated, narrow
structures rather than extended patches. In the inset to Figure~2, we plot
the direction of $\nabla \mathbi{P}$ for a small subregion of the image,

\affiliations{}

\noindent demonstrating that $\nabla \mathbi{P}$ changes most rapidly along
directions oriented perpendicular to the elongation of the filaments.  We
can explore the frequency dependence of these filaments, because the 1.4-GHz
ATCA data shown in Figure~1 consist of nine independent spectral channels of
width 8~MHz, spread over a total bandwidth of 96~MHz.  We have constructed
images of $|\nabla \mathbi{P}|$ for each individual spectral channel, and
these show the same set of specific features as in the overall image, albeit
at reduced signal-to-noise ratios. The lack of frequency dependence
indicates that the high-gradient structures seen in this data set correspond
to physical features in the ISM rather than to contour lines introduced by
the particular combination of observing frequency and angular resolution
used.\cite{fs07,new09}

\begin{figure*}
\includegraphics[width=\textwidth]{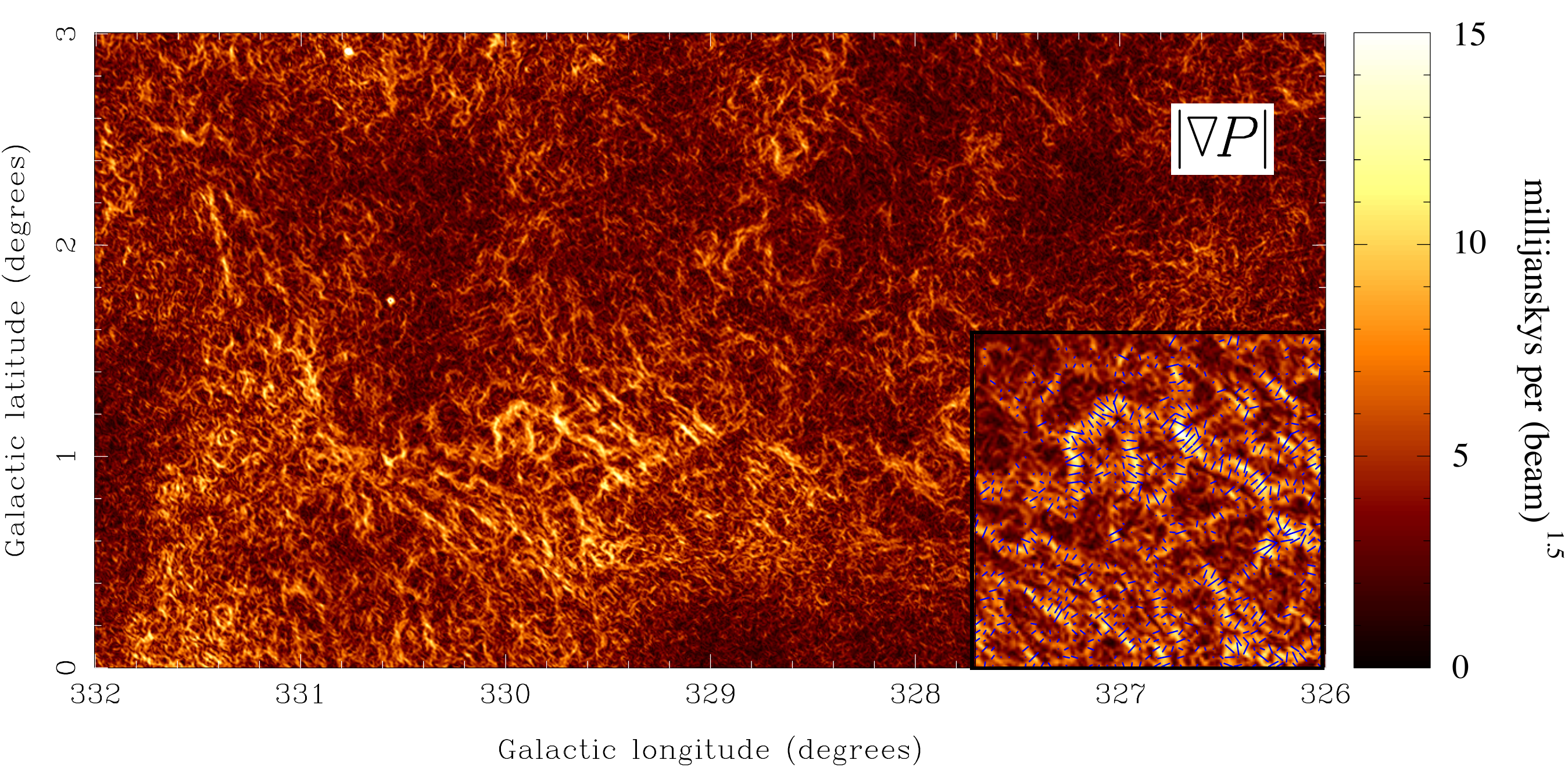}
\caption{$|$ {\bf{The gradient image of linear polarization, $|\nabla
\mathbi{P}|$, for an 18-deg$^2$ region of the Southern Galactic Plane
Survey.}}  $|\nabla \mathbi{P}|$ has been derived by applying
Equation~(\ref{eqn_gradp}) to the $Q$ and $U$ images from Fig.~1;  note
that $|\nabla \mathbi{P}|$ cannot be constructed from the scalar quantity $P
\equiv (Q^2 + U^2)^{1/2}$, but is derived from the vector field $\mathbi{P}
\equiv (Q,U)$.  
$|\nabla
\mathbi{P}|$ is a gradient in one dimension, for which the appropriate
units are (beam)$^{-0.5}$. Because $P$ measures
linearly polarized intensity in units of millijanskys per beam, $|\nabla
\mathbi{P}|$ has units of millijanskys per (beam)$^{1.5}$.  The 
scale showing $|\nabla
\mathbi{P}|$ 
is shown to the right of the image, and ranges from 0 to
15~mJy per (beam)$^{1.5}$. The inset shows an expanded 
version of the  structure with highest $|\nabla \mathbi{P}|$,
covering a box of side $0\fdg9$ centred on Galactic longitude $329\fdg8$ and
Galactic latitude $+1\fdg0$. Plotted in the inset is the
direction of $\nabla \mathbi{P}$ at each position, defined as $\arg(\nabla
\mathbi{P}) \equiv \tan^{-1} \left[ {\rm sign}\left( \frac{\partial
Q}{\partial x} \frac{\partial Q}{\partial y} + \frac{\partial U}{\partial x}
\frac{\partial U}{\partial y} \right) {\sqrt{\left(\frac{\partial
Q}{\partial y}\right)^2 + \left(\frac{\partial U}{\partial y}\right)^2}} /
{\sqrt{\left(\frac{\partial Q}{\partial x}\right)^2 + \left(\frac{\partial
U}{\partial x}\right)^2}} \right]$.  For clarity, vectors are only shown at
points where the amplitude of the gradient is greater than 5~mJy per
(beam)$^{1.5}$.}
\label{fig_gradp}
\end{figure*}

We first consider the possibility that these filaments of high gradient are
intrinsic to the source of emission.  Abrupt spatial transitions in the
strength or geometry of the magnetic field in a synchrotron-emitting region
would generate a large gradient in $(Q, U)$.  However, processes of that
sort would also produce structure in the overall synchrotron emissivity, such 
that we would observe features in the image of Stokes $I$ that match those
seen in $|\nabla \mathbi{P}|$.  No such correspondence is observed,
demonstrating that the regions of high polarization gradient are not
intrinsic to the source of polarized emission but must be induced by
Faraday rotation in magneto-ionized gas.

Because the amount of Faraday rotation is proportional to the line integral of
$n_e B_\parallel$ from the source to the observer 
(where $n_e$ is the density of free electrons and
$B_\parallel$ is the uniform component of the line-of-sight magnetic field),
the filamentary structure seen in $|\nabla \mathbi{P}|$ must correspond to
boundaries across which $n_e$ and/or $B_\parallel$ show a sudden increase or
decrease over a small spatial interval.  Such discontinuities could be shock
fronts or ionization fronts from discrete sources, as have been observed in
polarization around the rims of supernova remnants, \HII\ regions and
planetary nebulae.\cite{gdm+01,rukl08} We have examined this possibility by
carefully comparing our image of $|\nabla \mathbi{P}|$ with images
and gradient images of Stokes~$I$ (tracing shock waves seen in synchrotron
emission),\cite{gdm+01} 21-cm \HI\ emission (tracing atomic hydrogen)\cite{mgd+01}
and 656.3-nm H$\alpha$ (tracing ionized hydrogen)\cite{gmrv01,ppp+05}
over the same field, but do not find any correspondences. 

We conclude that the features seen in $|\nabla \mathbi{P}|$ are a generic
component of diffuse, ionized gas in this direction in the sky.  To test this
hypothesis, we performed a series of three-dimensional isothermal
simulations of magnetohydrodynamic turbulence in the ISM, each with
different parameters for the sonic Mach number, defined as $\mathcal{M}_s
\equiv \langle |{\bf v}|/c_s \rangle$, where $\bf{v}$ is the local velocity,
$c_s$ is the sound speed, and the averaging (indicated by angle brackets) 
is done over the whole
simulation.  For each simulation, we propagated a uniform source of
polarized emission through the distribution of turbulent, magnetized gas.
The resultant Faraday rotation produces a complicated distribution on the
sky of Stokes~$Q$ and $U$, from which we generated a map of the
polarization gradient using Equation~(\ref{eqn_gradp}).  Images of $|\nabla
\mathbi{P}|$ for representative simulations of the subsonic, transonic and
supersonic regimes are shown in Figure~3.  Narrow, elongated filaments of
high polarization gradient are apparent in each simulation in Figure~3,
although they differ in their morphology and degree of organization.  In
particular, the supersonic case (Fig.~3c) shows localized groupings of very
high-gradient filaments, corresponding to ensembles of intersecting
shocks.\cite{knpw07,blc05,ls09} By contrast, the subsonic (Fig.~3a) 
and transonic (Fig.~3b) cases show more-diffuse networks of filaments,
representing the cusps and discontinuities characteristic of any turbulent
velocity field.\cite{klb07,blc05,eyi95}

\begin{figure*}
\includegraphics[width=\textwidth]{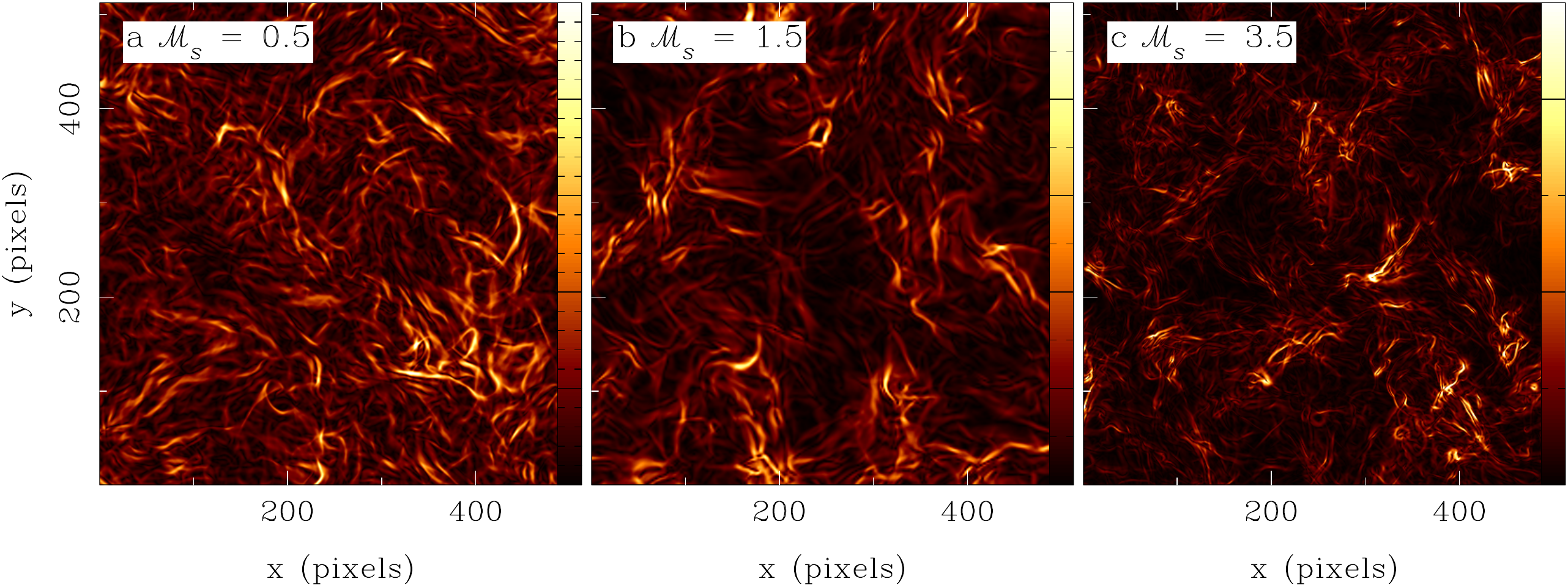}
\caption{{\bf {The gradient image, $|\nabla \mathbi{P}|$, derived from
propagation of linear radio polarization through three different isothermal
simulations of magnetized turbulence.}}  Each simulation is a
$512\times512\times512$-element periodic box with a linear dimension of
0.15~parsecs for each pixel, evolved in time using an essentially
nonoscillatory scheme.\protect\cite{bfkl09,cl03b} Three such simulations are
shown, each labelled with its corresponding value for $\mathcal{M}_s$ ---
subsonic ($\mathcal{M}_s < 1$; a), transonic
($\mathcal{M}_s \sim 1$; b) and supersonic cases ($\mathcal{M}_s
> 1$; c).  At the start of each simulation, the electron density
had a uniform value $n_e = 0.1$~cm$^{-3}$ and the magnetic field was
oriented in the plane of the sky with a uniform amplitude of $B = 0.3$~$\mu$G
(subsonic), 1~$\mu$G (transonic) or 2~$\mu$G (supersonic), which corresponds
to a constant Alfv\'enic Mach number of $\mathcal{M}_A = 2$ in each case.
Turbulence was driven solenoidally in Fourier space  at large scales (small
wavenumber) until the turbulent cascade had fully developed and a steady
state between input energy and dissipation had been reached.  In each case,
we illuminated the simulation volume with a background radio source of
uniform polarization at an emission frequency of 1.4~GHz, with $Q =
100$~mJy per pixel and $U = 0$~mJy per pixel at all
positions.  At each pixel, the line integral of $n_e B_\parallel$ was
computed, and the corresponding Faraday rotation applied to the polarized
signal, to calculate values of $Q$ and~$U$. No effects of finite
angular resolution, depolarization or incomplete interferometric visibility
coverage were included, so that the observed polarized signal is $P =
100$~mJy per pixel at all positions.  We then calculated the
gradient, $|\nabla \mathbi{P}|$, using Equation~(\ref{eqn_gradp}).  The
scales showing $|\nabla \mathbi{P}|$
are shown on the right of the images, and range
from 0 to 25 mJy per (pixel)$^{1.5}$ for panel a, 0 to 100
mJy per (pixel)$^{1.5}$ for panel b, and 0 to 500 mJy per
(pixel)$^{1.5}$ for panel c.}
\label{fig_sim}
\end{figure*}

Visual comparison of the simulated distributions of  $|\nabla \mathbi{P}|$
with real data  (Fig.~2) suggests that the  subsonic and transonic cases
shown in Figures~3a,b more closely resemble the observations than does the
supersonic case.  We can quantify this statement by calculating the
third-order moment (skew, $\gamma$) and the fourth-order (kurtosis, $\beta$)
moments of the probability distribution function of $|\nabla
\mathbi{P}|$ for both observations and simulations: these quantities
parameterize the degree of Gaussian asymmetry in the probability
distribution function, and hence provide
information on the level of compression due to shocks in the
data.\cite{klb07,bfkl09}

In the simulations, we found that both the skew and kurtosis of $|\nabla
\mathbi{P}|$ increased monotonically with sonic Mach number.  We used a
genetic algorithm\cite{whi94b} to determine that the threshold for strongly
supersonic turbulence was $\gamma > 1.1$ and $\beta > 1.5$.  We then
computed the third- and fourth-order moments for the observed distribution
of $|\nabla \mathbi{P}|$ shown in Figure~2, and found that $\gamma = 0.3$
and $\beta = 0.9$.  

This analysis of the moments of the polarization gradient therefore
confirms quantitatively what we concluded above from visual inspection: the
turbulent, ionized ISM in this direction is subsonic or transonic.  The
findings we obtained by imaging the polarization gradients produced by
interstellar turbulence are supported by recent statistical studies of
H$\alpha$ emission measures and of 21-cm \HI\ column densities over
large volumes, which have similarly found $\mathcal{M}_s \la 2$ for warm gas
throughout the ISM.\cite{hbk+08,bslk10}

In the simulations shown in Figure~3, the sharp gradients in $(Q,U)$ occur
as a result of localized high values of the gas density and magnetic field,
resulting from vorticity or shock compression.  However, the filamentary
features seen in  $|\nabla \mathbi{P}|$ may not be easily observable in
other types of data: for example, if we adopt typical parameters for warm
ionized gas\cite{fer01,gmcm08} of $n_e \approx 0.3$~cm$^{-3}$ and
$B_\parallel \approx 2$~$\mu$G, even the compression associated with a
strong adiabatic shock produces across-filament 
changes in emission measure and Faraday
rotation measure  of only $\approx0.5$~parsecs~cm$^{-6}$ and
$\la5$~radians~m$^{-2}$, respectively, assuming a spatial
scale\cite{kkfg08,hh04b} for these structures of $\sim 0.5$~parsecs.
This is below observable levels in H$\alpha$ and other tracers of
emission measure.  The rotation measure gradient\cite{hh04b}  across these
interfaces is potentially observable in spectropolarimetric radio data, but
the addition of single-dish observations is required to recover the total power
of the polarized signal.  By contrast, even a small gradient in rotation
measure can produce an arbitrarily large value of $|\nabla \mathbi{P}|$
(irrespective of whether single-dish measurements are present in the data),
provided that there is a strong source of background polarized emission through
which the discontinuities in Faraday rotation are viewed.  Further
investigation of the polarization gradient and its statistical properties
will provide robust estimates of poorly constrained parameters of turbulent
flows such as the sonic and Alfv\'enic Mach numbers, the characteristic
magnetic field strength, the Reynolds number and the physical scale of
energy injection.

\vspace*{5mm}
\noindent
{\sf{\textbf{Received 3 March 2011; accepted 11 August 2011. \\
Published online 5 October 2011.}}}
\vspace*{-12mm}

\bibliographystyle{nature_djm}
\bibliography{journals_nature,modrefs,psrrefs}

\balance

\medskip
\vspace*{2mm}
\noindent
{\sf{\textbf{Acknowledgements}}
We thank S.~Brown, A.~Hill, R.~Kissmann, A.~MacFadyen, M.-M.~Mac~Low,
E.~Petroff, P.~Slane and X.~Sun for useful discussions, and the anonymous
referees for insightful recommendations.  The Australia Telescope Compact
Array is funded by the Commonwealth of Australia for operation as a National
Facility managed by CSIRO.  B.M.G.\ and T.R.\ acknowledge the support of the
Australian Research Council through grants FF0561298, FL100100114 and
FS100100033.  B.B. acknowledges support from the National Science Foundation
Graduate Research Fellowship and the NASA Wisconsin Space Grant Institution.
A.L. acknowledges the support of the National Science Foundation through
grant AST0808118 and of the Center for Magnetic Self-Organization in
Astrophysical and Laboratory Plasmas.  We thank the staff of the Australia
Telescope National Facility for their support of the Southern Galactic Plane
Survey, especially M.~Calabretta, R.~Haynes, D.~McConnell, J.~Reynolds,
R.~Sault, R.~Wark and M.~Wieringa.  }

\medskip
\vspace*{2mm}
\noindent
{\sf{\textbf{Author Contributions}}
J.M.D., N.M.Mc-G., B.M.G. and A.J.G.  carried out the original observations.
B.M.G., N.M.Mc-G. and T.R. produced the polarization images from the raw
data.  B.M.G., M.H., K.J.N-Mc., R.D.E and N.M.Mc-G. worked together to
develop the gradient technique, and B.M.G. then applied the gradient
technique to the images. B.B. and A.L. performed the simulations and
statistical analysis.  B.M.G.  led the writing of the paper and
interpretation of results.  All authors discussed the results and commented
on the manuscript.
}

\medskip
\vspace*{2mm}
\noindent
{\sf{\textbf{Author Information}} 
Reprints and permissions information is available at
www.nature.com/reprints. The authors declare no competing financial
interests.  Readers are welcome to comment on the online version of this
article at www.nature.com/nature.  Correspondence and requests for materials
should be addressed to B.M.G.  (bryan.gaensler@sydney.edu.au).  
}

\clearpage

\label{lastpage}
\end{document}